\documentclass[a4paper,11pt]{article}
\pdfoutput=1 

\usepackage{jheppub} 

\usepackage[T1]{fontenc} 
\usepackage{amssymb}
\usepackage{amsmath}
\usepackage{color}
\usepackage{xspace}
\usepackage{appendix}
\usepackage{float}
\usepackage{hyperref}
\usepackage{diagbox}
\usepackage[utf8x]{inputenc}
\graphicspath{{figures/}}

\title{  Transverse-Energy-Energy Correlations in Deep Inelastic Scattering}
\author[a]{Hai Tao Li}
\author[a]{, Ivan Vitev}
\author[b]{and Yu Jiao Zhu}

\affiliation[a]{Los Alamos National Laboratory, Theoretical Division, Los Alamos, NM, 87545, USA}
\affiliation[b]{Zhejiang Institute of Modern Physics, Department of Physics, Zhejiang University, Hangzhou, 310027, China}

\emailAdd{haitaoli@lanl.gov}
\emailAdd{ivitev@lanl.gov}
\emailAdd{zhuyujiao@zju.edu.cn}

\abstract{
Event shape observables have been widely used for precision QCD studies at various lepton and hadron colliders. We present the most accurate calculation of the transverse-energy-energy correlation event shape variable in deep-inelastic scattering. In the framework of soft-collinear effective theory the cross section is factorized as the convolution of the hard function, beam function, jet function and soft function in the back-to-back limit. A close connection to  TMD factorization is established,  as the beam function when combined with part of the soft function is identical to the  conventional TMD parton distribution function,  and the jet function is the second moment of the TMD fragmentation function matching coefficient.  We validate our  framework by comparing the obtained LO and NLO leading singular distributions to the full QCD calculations in the back-to-back limit.  We report  the resummed  transverse-energy-energy correlation distributions up to N$^3$LL accuracy matched with the NLO cross section for the production of a lepton and two jets. Our work provides a new way to precisely study TMD physics at the future Electron-Ion Collider. 
}

\begin{document} 
\maketitle
\flushbottom

\section{Introduction}

Event shape observables  describe the  patterns,  correlations, and energy flow of hadronic final states in high energy processes. They have been widely investigated to study the various dynamical  aspects  of  QCD  in $e^+e^-$, $ep$,  $pp$, and heavy-ion collisions. Event shape variables can be used to determine the strong coupling $\alpha_s$ and test asymptotic freedom, to tune the nonperturbative  Quantum Chromodynamics (QCD)  power  corrections, and to search for new physics phenomena.  Furthermore,  these observables can be studied with high precision theoretically  and compared to experimental measurements  at the future Electron-Ion Collider~(EIC).  

There are many efforts devoted to  the study of event shape observables in deep-inelastic scattering (DIS).  The next-to-leading order (NLO) QCD corrections were obtained about twenty years ago~\cite{Catani:1996vz,Graudenz:1997gv,Nagy:2001xb}.  Recently, the  next-to-next-to-leading order~(NNLO)  QCD  corrections to various event shape distributions were computed in ref.~\cite{Gehrmann:2019hwf}. Near the infrared region resummation is required to obtain  reliable predictions, which are available at next-to-leading logarithmic~(NLL) level~\cite{Antonelli:1999kx, Dasgupta:2001eq,Dasgupta:2002dc,Dasgupta:2003iq} for most event shape observables, next-to-next-to-leading logarithmic~(NNLL) level for 1-jettiness~\cite{Kang:2013lga} and  angularity~\cite{Kang:2019bpl},  and  next-to-next-to-next-to-leading logarithmic~(N$^3$LL) level for thrust~\cite{Kang:2015swk}. On the experimental side, H1 and ZEUS collaborations have measured some event shape variables at  HERA ~\cite{Adloff:1997gq,Breitweg:1997ug, Adloff:1999gn,Chekanov:2002xk, Aktas:2005tz,Chekanov:2006hv}. With more precise measurements in DIS at the future EIC, event shape observables can serve as a  precision test of QCD and new probes to reveal the proton or nuclear structure. 

Here, we will concentrate on the transverse-energy-energy correlation (TEEC) event shape observable in DIS. TEEC~\cite{Ali:1984yp} at hadronic colliders is an extension of the energy-energy correlation (EEC)~\cite{Basham:1978bw} variable  introduced decades ago in $e^+e^-$ collisions to describe the global event shape. The EEC is defined as 
\begin{align}
    \text{EEC} = \sum_{a,b} \int d\sigma_{e^+e^-\to a+b+X} \frac{2 E_{a} E_b}{|\sum_{i}E_{i}|^2 }\delta(\cos\theta_{ab}-\cos\theta) \, , 
\end{align}
where $E_{i}$ is the energy of hadron $i$ and $\theta_{ab}$ is the opening angle between hadrons $a$ and $b$. New studies of EECs, which include  analytical NLO calculations~\cite{Dixon:2018qgp,Luo:2019nig}, NNLO in $\mathcal{N}=4$ super-Yang-Mills theory~\cite{Belitsky:2013xxa,Belitsky:2013bja,Belitsky:2013ofa,Henn:2019gkr}, all-order factorization in QCD in the back-to-back limit~\cite{Moult:2018jzp}, and all order structure in the collinear limit~\cite{Dixon:2019uzg,Kologlu:2019mfz,Korchemsky:2019nzm,Chen:2020uvt}, have furthered our understanding of this observable.

At hadronic colliders, where detectors lack the typical hermeticity of detectors at $e^+e^-$ machines, the event shape observable can be generalized by considering the transverse energy of the hadrons. TEEC, which is defined as 
\begin{align}
    \text{TEEC} = \sum_{a,b} \int d\sigma_{pp\to a+b+X} \frac{2 E_{T,a} E_{T,b}}{|\sum_{i}E_{T,i}|^2} \delta(\cos\phi_{ab}-\cos\phi)\, , 
    \label{teec}
\end{align}
was investigated  in refs.~\cite{Ali:2012rn,Gao:2019ojf, Gao:2019prep}.  In eq.~(\ref{teec})  $E_{T,i}$ is the transverse energy of hadron $i$ and $\phi_{ab}$ is the azimuthal angle between hadrons $a$ and $b$. The   NLO QCD corrections for the TEEC observable were calculated in ref.~\cite{Ali:2012rn}.  The works~\cite{Gao:2019ojf,Gao:2019prep} investigated TEEC in the dijet limit and showed that it exhibits remarkable perturbative simplicity.  

 In DIS, TEEC can be generalized by considering the transverse-energy and transverse-energy correlation between the lepton and hadrons in the final state,  which is first studied in this work. 
 We define this event shape observable as follows:
\begin{align}~\label{eq:teec_dis}
    \text{TEEC} =&  \sum_{a} \int d\sigma_{lp\to l+a+X} \frac{ E_{T,l}  E_{T,a}}{E_{T,l} \sum_{i} E_{T,i}}  \delta(\cos\phi_{la}-\cos\phi) 
    \nonumber \\ 
    =& \sum_{a} \int d\sigma_{lp\to l+a+X} \frac{  E_{T,a}}{\sum_{i} E_{T,i}}  \delta(\cos\phi_{la}-\cos\phi) \, ,
\end{align}
where the sum runs over all the hadrons in the final states and $\phi_{la}$ is the  azimuthal angle between final state lepton $l$ and hadron $a$. Note that there is no QCD collinear singularity ($\phi_{la}\to 0$) along the outgoing lepton's momentum in DIS, for which one needs to perform  the  resummation  at hadron colliders and in $e^{+}e^{-}$ annihilation~\cite{Dixon:2019uzg}.  As we will show below, resummed predictions in the back-to-back limit ($\phi_{la}\to \pi$) can be obtained to high accuracy and the distribution in the whole range $\phi\in [0, \pi]$ can be reliably calculated. One of the advantages of EEC/TEEC is that the contribution from soft radiation is suppressed as it carries  parametrically small energy. Therefore, the hadronization effects are expected to be small in comparison to other event shape observables. TEEC in DIS can be used to determine the strong coupling precisely similar to analysis in refs.~\cite{Catani:1996vz,Graudenz:1997gv,Nagy:2001xb} and to study the nuclear dynamic as in ref.~\cite{Kang:2013lga}.  Additionally it is also feasible to study transverse-momentum dependent (TMD) physics using TEEC in DIS.

In this paper, we present our study of TEEC in the DIS process.  Similar to EEC~\cite{Moult:2018jzp} in $e^+e^-$ collisions and TEEC~\cite{Gao:2019ojf} in hadronic collisions, the cross section for this observable in the back-to-back limit can be factorized as the convolution of the hard function, beam function, jet function, and soft function using the frameworks of soft-collinear effective theory (SCET)~\cite{Bauer:2001yt, Bauer:2001ct, Bauer:2000yr, Bauer:2000ew,Beneke:2002ph}.  This approach is similar to a 1-dimension TMD factorization and is, thus, closely related to TMD physics. The beam functions are identical to the TMD parton distribution functions (PDFs) and the jet function is the second moment of the matching coefficients of the TMD fragmentation functions. For details see refs.~\cite{Moult:2018jzp,Luo:2019hmp,Luo:2019bmw}.   Furthermore, the factorization formalism is also similar to the usual TMD factorization, for example see the work~\cite{Gutierrez-Reyes:2019vbx} for N$^3$LL jet $q_T$ distribution.
The  non-trivial LO and NLO  QCD distributions of the TEEC  observable are reproduced by the leading power SCET in the back-to-back limit, which validates our formalism. Resummation can be achieved by evolving  each component of the factorized expression from its intrinsic scale to a  suitably chosen common scale.  The main goal of our work is to present the most precise TEEC  predictions in DIS at N$^3$LL+NLO. The effects of the nonperturbative physics are also discussed.   Since there is no collinear singularity, we are able to provide the N$^3$LL+NLO distribution in the complete range of $0<\phi<\pi$. 
The perturbative behavior of this observable is under  good theoretical control in QCD, which can be further improved if one can match the resummation with NNLO corrections.  Consequently, it can be used to study the non-perturbative physics in a precise quantitative manner.

The rest of this paper is organized as follows. In the next section the factorization formalism for the TEEC observable is present. The hard function, beam function, jet function, and  soft function are discussed. In Section~\ref{sec:num} we investigate the hadronization effect using {\sc\small{ Pythia8}}.  We further verify the factorization formula by comparing  the LO and NLO singular distribution with the full QCD ones. The N$^3$LL and N$^3$LL+NLO predictions are also present. Finally, we conclude in Section~\ref{sec:concl}.  The RG equations and anomalous dimensions are present in Appendix~\ref{app:ad}.

\section{Theoretical formalism}\label{sec:Theory}

The underlying partonic Born process considered in this work is 
\begin{align}
    e(k_1) + q(k_2) \to e(k_3) + q(k_4) .
\end{align}
The first order non-trivial contribution to TEEC begins from one order higher.  
In the back-to-back limit, the TEEC cross section is defined as
\begin{align}
    \frac{d\sigma}{d\cos\phi} \approx \sum_{h} \int d\sigma_{l N\to l+h+X}\times  \frac{p_T^h}{p_T}  \times  \delta(\cos\phi_{lh} - \cos\phi)~,
\end{align}
where $p_T$ is the transverse momentum of the outgoing lepton.  We define the momenta of the event in the $x-z$ plane, i.e. at LO the components of all the momenta along $y$-direction are zero.  In the back-to-back limit it is convenient to introduce the variable   $\tau= (1+\cos\phi)/2$, related to the non-zero momentum balance along $y$-direction of the event due to soft and/or collinear radiations.  It can be written as 
\begin{align}
    \tau = \frac{\left|k_{2,y}-k_{s,y}+\frac{k_{4,y}}{\xi_4}\right|^2}{4p_T^{2}},
\end{align}
where $k_{s,y}$ is the $y$-momentum of the soft radiation and  $\xi_4$ is the momentum fraction of the hadron relative to the jet.  The soft radiation contributes through the recoil to the energetic collinear partons.
Similar to the case of EEC in electron-position collisions or TEEC at hadronic collisions, the cross section in the back-to-back limit is factorized into the convolution of a hard function, beam function, soft function, and jet function. \
Specifically, up to leading power in SCET the cross section can be written as  
\begin{align}\label{eq:sing}
    \frac{d\sigma^{(0)}}{d\tau} =& \sum_{f}  \int\frac{d\xi  dQ^2 }{\xi Q^2}  Q_{f}^2 \sigma_0 \frac{p_T}{\sqrt{\tau}}\int \frac{db}{2\pi} e^{-2ib\sqrt{\tau} p_T}  B_{f/N}(b,E_2, \xi, \mu, \nu ) H(Q, \mu )
    \nonumber \\ & \times 
    S\left(b,\frac{n_2\cdot n_4}{2},\mu,\nu\right)J_{f}(b,E_4, \mu, \nu) \,,
\end{align}
where $\sigma_0= \frac{2 \pi \alpha^2}{Q^2}[1+(1-y)^2] $,  $b$ is the conjugate variable to $k_y$,  $Q^2$ is the invariant mass of the virtual photon, and $y=Q^2/\xi/s$.  Four-vectors  $n_2$ and $n_4$ represent the momentum directions of the momenta $k_2$ and $k_4$, respectively.  $E_2$ and $E_4$ are the energies of $k_2$ and $k_4$. $\nu$ is rapidity scale associated with the rapidity regulator for which we adopt the exponential regulator  introduced in ref.~\cite{Li:2016axz}.

$B_{f/N}$, which describe the contribution from collinear radiation in the initial state, are the same as the usual TMD beam functions.
The operator definition for the beam function in SCET is 
\begin{align}\label{eq:beam}
    B_{q / N}\left( b,  \xi\right)\equiv & \int \frac{d b}{4 \pi} e^{-i \xi b P^{+} / 2} 
\left\langle N(P)\left|\bar{\chi}_{n}\left(0, b^{-}, b_{\perp}\right) \frac{\slash\!\!\!\bar{n}}{2} \chi_{n}(0)\right| N(P)\right\rangle\,,
\end{align}
with $\chi_n =W_n^\dagger \xi_n$, where $ \xi_n$ is the collinear quark field and $W_n$ is the path-ordered collinear Wilson line $W_n(x) = \mathcal{P} \exp \left(i g \int_{-\infty}^{0} d s \bar{n} \cdot A_{n}(x+\bar{n} s)\right)$.  In the operator definition, we suppress the arguments of  kinematics and scales.   
The TMD beam functions have been calculated up to three loops for quark beam functions and two loops for gluon beam functions~\cite{Gehrmann:2012ze, Gehrmann:2014yya,Luebbert:2016itl,Echevarria:2016scs,Luo:2019hmp, Luo:2019bmw,Luo:2019szz}.

The jet functions $J_f$ are defined as the second Mellin  moment of the matching coefficients of the TMD fragmentation function~\cite{Echevarria:2016scs, Echevarria:2015usa, Luo:2019hmp, Luo:2019bmw}. The explicit expression up to two loops for the jet functions can be found in refs.~\cite{Luo:2019hmp, Luo:2019bmw}.  

The operator definition for the soft function is 
\begin{align}
S_{\rm DIS}\left(b\right)
\equiv
  \frac{1}{N_{c}} \operatorname{Tr}
  \left\langle 0\left|
  \bar T\left[
  Y_{n_2}(0)Y^\dagger_{n_4}(0)
  \right]^\dagger
  \,
  T\left[
  Y_{n_2}(0)Y^\dagger_{n_4}(0)
  \right]
  \right| 0\right\rangle\,,
\end{align}
where $Y_{n_2}$ and $Y^\dagger_{n_4}$ correspond to an incoming quark and an outgoing quark, respectively. The explicit expressions of  $Y_{n_2}$ and $Y^\dagger_{n_4}$ are 
\begin{align}
Y_{n_2}(x)= \mathcal{P} \exp \left(i g_{s} \int_{-\infty}^{0} d s {n_2} \cdot A_{s}(x+s  {n_2})\right)\,,
 \nonumber \\
Y^\dagger_{n_4}(x)= \mathcal{P} \exp \left(i g_{s} \int_{0}^{\infty} d s {n_4} \cdot A_{s}(x+s  {n_4})\right)\,.
\end{align} 
We suppress the arguments of  kinematics and scales in the operator definition.    The soft function for TEEC in DIS can be written in terms of the soft function in EEC in $e^+e^-$ collisions, which can be written as  
\begin{align}
    S\left(b, \frac{n_2\cdot n_4}{2},\mu, \nu \right) =  S_{\rm EEC}\left(L_b,L_\nu+\ln \frac{n_2 \cdot n_4}{2} \right) \, , 
\end{align}
where $S_{\rm EEC}$ is the soft function for EEC. In the above $L_\nu = \ln \nu^2 b^2/b_0^2$ and $L_b=\ln\mu^2 b^2/b_0^2$ with $b_0=2 e^{-\gamma_{\rm E}}$. $S_{\rm EEC}$ is identical to TMD soft function~\cite{Moult:2018jzp}. Up to three loops the expression for the soft function can be found in refs.~\cite{Li:2016ctv}. 

The hard function encodes the short-distance physics, which is the matching coefficient from full QCD onto SCET. The analytical expression of $H( Q, \mu)$ up to NNLO is given in ref.~\cite{Becher:2006mr} and the one at three-loop level  can be obtained from the quark form factor, as shown in refs.~\cite{Baikov:2009bg,Gehrmann:2010ue}.

The renormalization group~(RG) equations and anomalous dimensions  needed for our calculation are given in  Appendix~\ref{app:ad}.
With all the components and their RG equations available, we can achieve precision predictions for this observable up to N$^3$LL.  The resummed cross section is obtained by evolving the hard function from $\mu_h$ to $\mu_c$ and the soft function from $(\mu_s, \nu_s)$ to $(\mu_c, \nu_c)$. It can be written as 
\begin{align}{\label{eq:resum}}
    \frac{d\sigma^{(0)}_{\rm RES }}{d\tau }=&  \sum_{f}  \int\frac{d\xi  dQ^2 }{\xi Q^2}  Q_{f}^2 \sigma_0 \frac{p_T}{\sqrt{\tau}} \int \frac{db}{2\pi} e^{-2ib\sqrt{\tau} p_T}  B_{f/N}(b,E_2, \xi, \mu_c, \nu_c ) H( Q, \mu_h )
    \nonumber \\ & 
   \times  S\left(b, \frac{n_2\cdot n_4}{2}, \mu_s,\nu_s \right)J_{f}(E_4, b, \mu_c, \nu_c) \exp\left[ \int_{\mu_h}^{\mu_c} \frac{d\bar{\mu}}{\bar{\mu}} \Gamma_h(\bar{\mu}) +\int^{\mu_c}_{\mu_s} \frac{d\bar{\mu}}{\bar{\mu}} \Gamma_s(\bar{\mu}, \nu_s)  \right] \
   \nonumber \\ &  \times 
   \exp\left[\int_{\nu_s}^{\nu} \frac{d\bar{\nu}}{\bar{\nu}} \Gamma_r(\mu_c, \mu_b) \right]  \, , 
\end{align}
where $\Gamma_h$ and $\Gamma_s$ are the anomalous dimensions of the hard and soft functions, and $\Gamma_r$ is the rapidity anomalous dimension of the soft function. 

The prediction away from the back-to-back limit is obtained through matching the resummed calculations with the fixed-order ones, which can be written as 
\begin{align}
    \frac{d\sigma_{\rm N^{l}LL+N^{k}LO}}{d\tau}=  \frac{d\sigma_{\rm N^{l}LL}}{d\tau} +  \frac{d\sigma_{\rm N^{k}LO}}{d\tau} - \left( \frac{d\sigma_{\rm N^{k}LO}}{d\tau}\right)_{\rm sing.}~. 
\end{align}
The singular distribution $\left( \frac{d\sigma_{\rm N^{k}LO}}{d\tau}\right)_{\rm sing.}$ is the fixed-order prediction from eq.~(\ref{eq:sing}) in the leading power of SCET, which captures the singular behavior of the  QCD fixed-order predictions  in the leading power in the back-to-back limit.

\section{Numerical results}~\label{sec:num}

We will present  numerical predictions with enter-of-mass energy  $\sqrt{s}=141$ GeV, corresponding to beam energies 20 (lepton) GeV$\times$250 (proton) GeV,  typical  for the future EIC~\cite{Aschenauer:2014cki}.  We also consider enter-of-mass energy  $\sqrt{s}=318$ GeV, corresponding to beam energies 27.5 GeV$\times$920 GeV at HERA.  We select events with  constraints on the transverse momentum of the outgoing lepton $p_T^{l}> 20$ GeV and  $p_T^{l}> 30$ GeV for 141  GeV and 318 GeV electron-proton collisions, respectively. All  calculations are performed using PDF4LHC15\_nnlo\_mc PDF sets~\cite{Butterworth:2015oua,Dulat:2015mca,Harland-Lang:2014zoa,Ball:2014uwa}  and the associated strong coupling provided by  {\sc \small Lhapdf6}~\cite{Buckley:2014ana}.

\subsection{{\sc\small Pythia} Simulation}
\begin{figure}
    \centering
    \includegraphics[width=0.49 \textwidth]{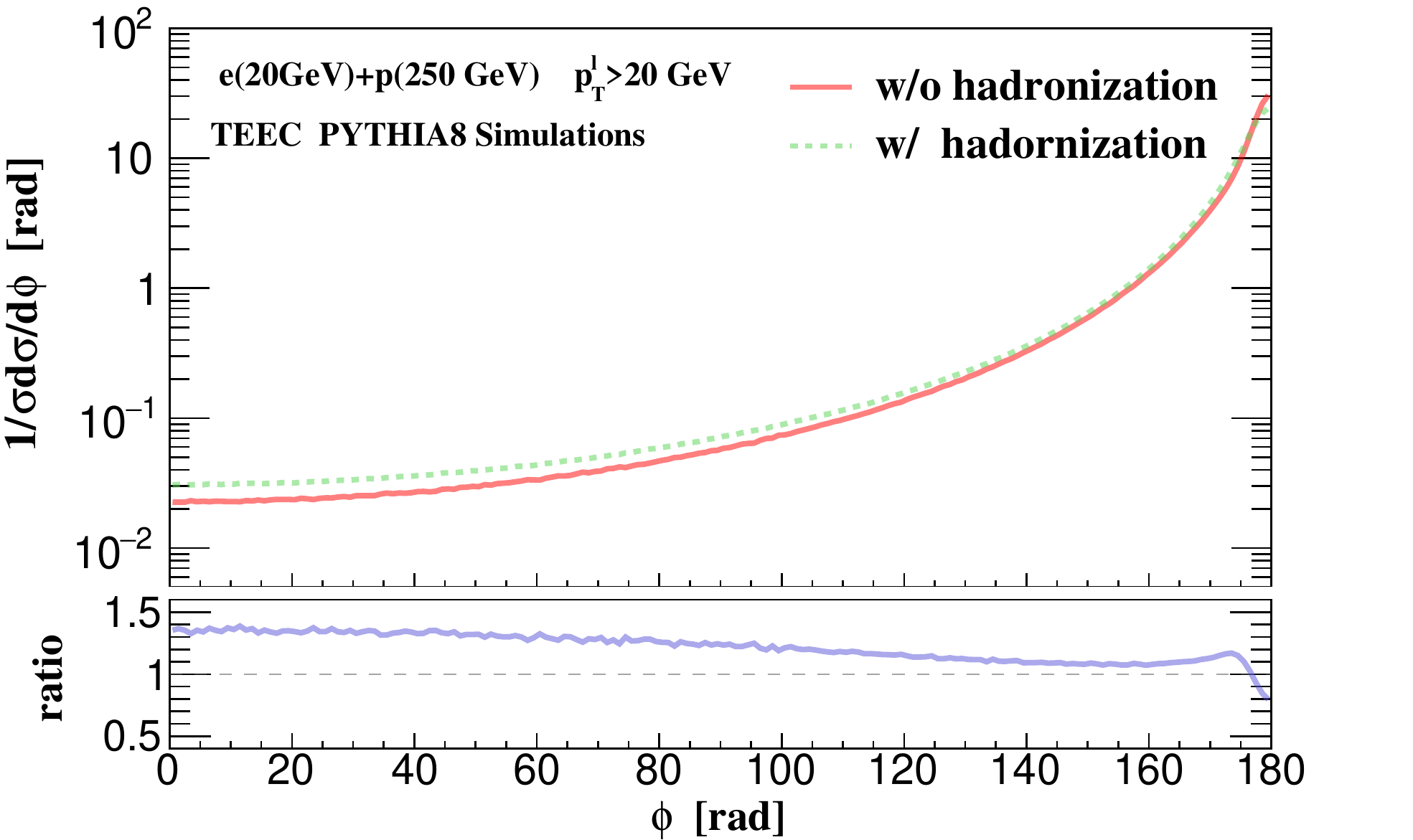}
    \includegraphics[width=0.49 \textwidth]{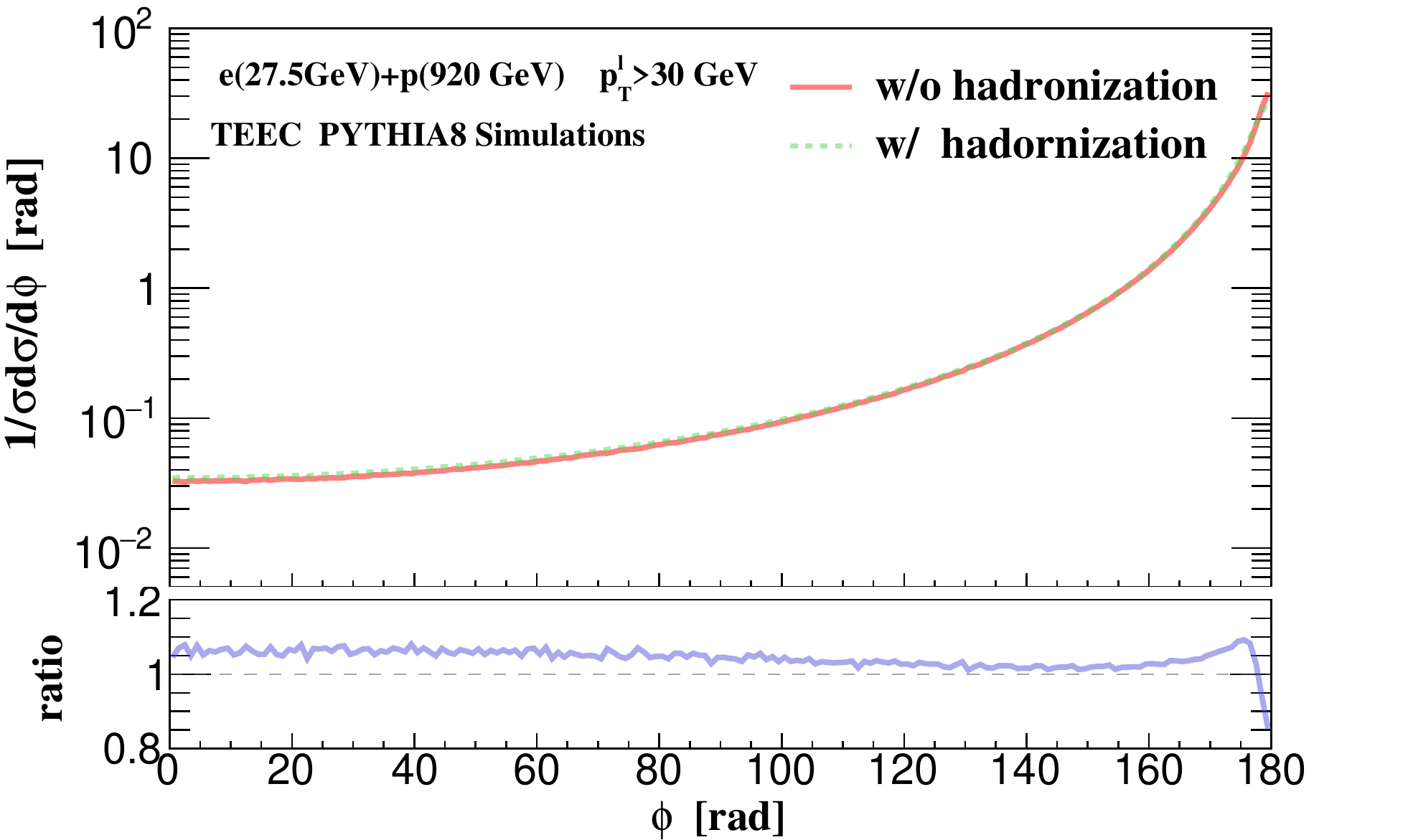}
    \caption{{\sc \small Pythia} simulations of the TEEC distribution versus $\phi$ with and without hadronization effects. We consider center-of-mass energy  $\sqrt{s}=141$ GeV with $p_T^{l}>20$ GeV (left),  and enter-of-mass energy  $\sqrt{s}=318$ GeV with $p_T^{l}>30$ GeV (right).  The ratio is defined as predictions with hadronization effects divided by the ones without hadronization effects.    }
    \label{fig:pythia}
\end{figure}

To assess the effects of hadronization  on the TEEC observable we start with  {\sc \small Pythia8}~\cite{Sjostrand:2007gs,Sjostrand:2014zea} simulations.
Figure~\ref{fig:pythia} shows the predictions for normalized TEEC with and without hadronization in $\sqrt{s}=$141 Gev (left) and $\sqrt{s}=$318 GeV (right) $ep$ collisions.  The lepton is selected with a finite transverse momentum in the final state and there is no divergence in the usual collinear limit ($\phi\to 0 ^o$). The cross section is dominated by the back-to-back region, where there are collinear and/or soft singularities in fixed-order calculations.
As shown in Fig.~\ref{fig:pythia}, the hadronization effects are more important in  $\sqrt{s}=141$~GeV $ep$ collisions since  the tagged lepton is of a smaller $p_T$.  The corrections themselves  are about 20 to 35 percent for small  and large $\phi$.  For   $\sqrt{s}=318$~ GeV collisions the hadronization effects are a few percent for small $\phi$ and  about 15\% in the back-to-back region. 
In comparison to other event shape observables, as can be seen in the simulations in ref.~\cite{Gehrmann:2019hwf}, the overall hadronization effects are much smaller. The reason behind this observation is that the soft particle contribution is suppressed by the energy  in  the TEEC. Therefore, the predictions for the TEEC observable can be significantly improved through  high order calculations in  perturbative QCD.
In this section we will show that N$^3$LL TEEC in the back-back limit is under very good control after resummation and the nonperturbative effects are investigated. In the future the distributions can be further improved with NNLO calculations.
Subsequently, the observable can be used  to test perturbative QCD and measure the QCD coupling in a unique way.

\subsection{Fixed-order results}

\begin{figure}
    \centering
    \includegraphics[width=0.49 \textwidth]{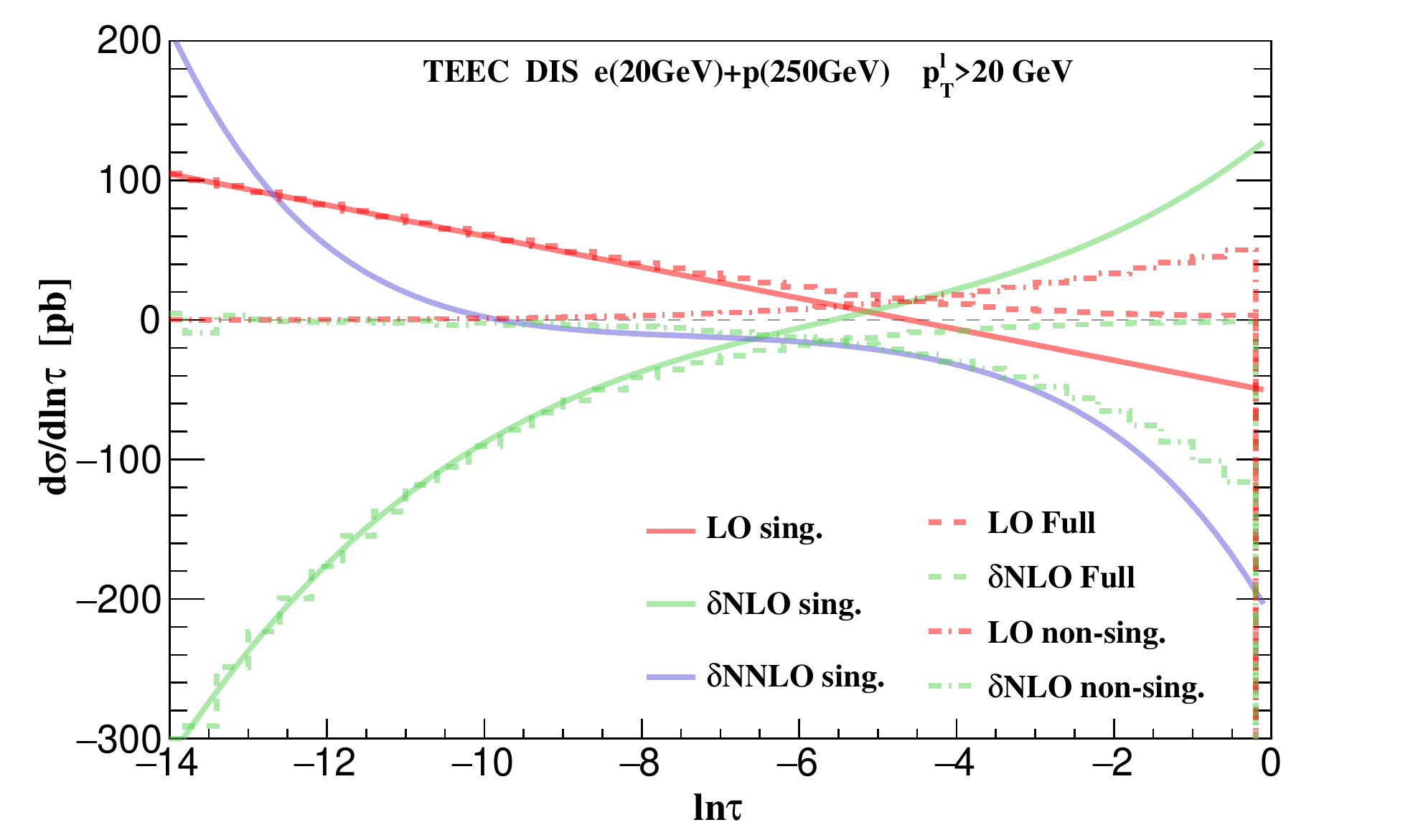}
    \includegraphics[width=0.49 \textwidth]{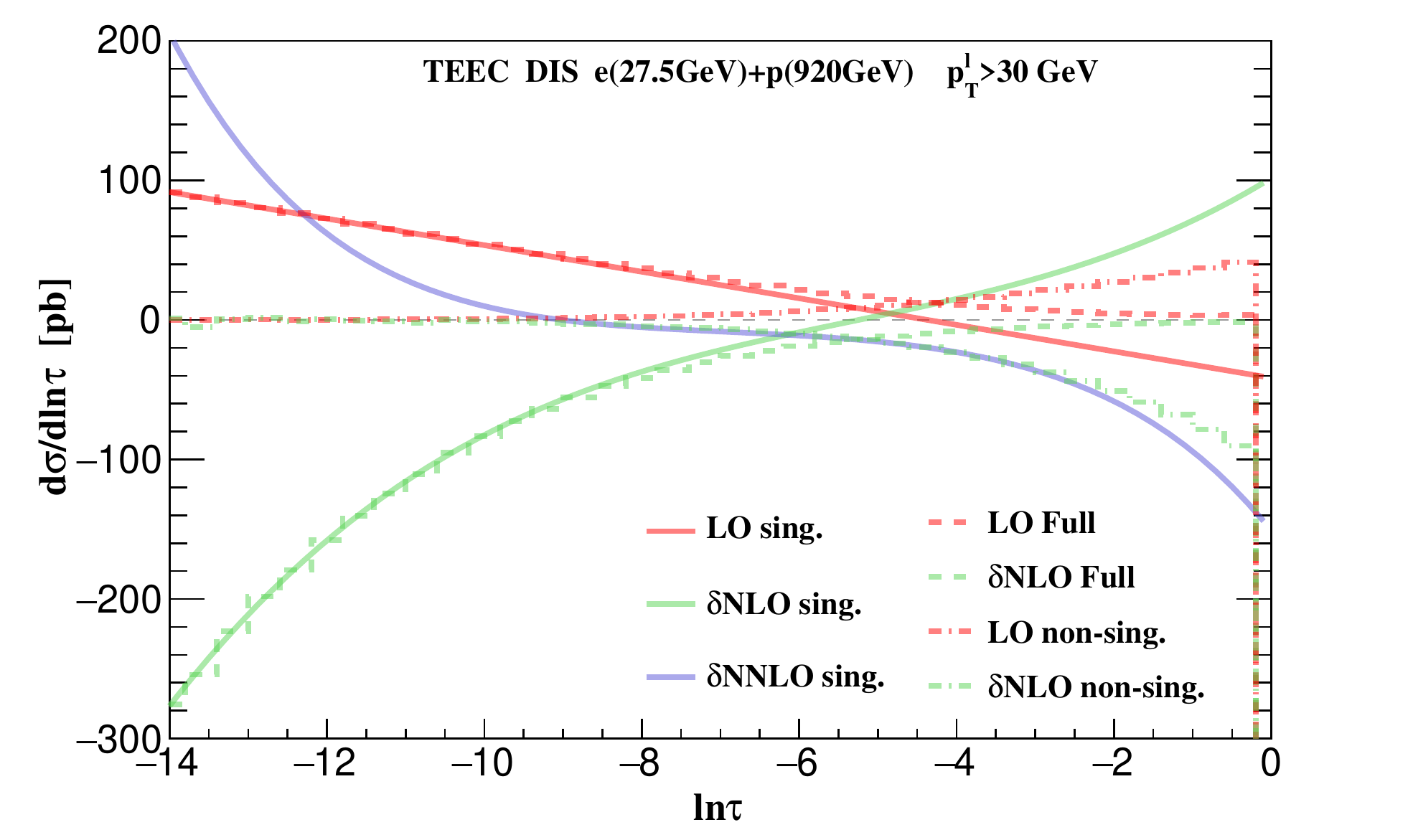}
    \caption{Fixed-order results for the  $\ln\tau$ distributions in the back-to-back limit for $\sqrt{s}=$ 141 GeV (left) and 318 GeV(right) $ep$ collisions. The full QCD (dash lines) and non-singular (dash-dotted lines) distributions  are shown up to NLO, while the leading singular (solid lines) distributions are up to NNLO.  }
    \label{fig:fixed-order}
\end{figure}

We now move on to the core calculations in our work on TEEC in DIS. In Fig.~\ref{fig:fixed-order}
we present the comparison of the leading singular distributions from SCET to full QCD calculations. 
Figure~\ref{fig:fixed-order} also shows the non-singular contributions which are defined as the differences between the full QCD  and the singular calculations.
With the known components in eq.~(\ref{eq:sing}) up to two loops, the LO and $\delta$NLO singular distributions are present with solid orange and solid green  lines. With three-loop anomalous dimensions the singular distributions are calculated up to NNLO in QCD and are shown as the solid blue lines. The singular distributions oscillate between $\infty$ and $-\infty$ from LO to NNLO when $\tau \to 0$.  The full LO and NLO results for two jet production in DIS are calculated making use of {\sc \small Nlojet++}~\cite{Nagy:2001xb,Nagy:2005gn} and are denoted by dashed lines. Finally, the dash-dotted lines stand for the non-singular distributions.   The renormalization  and factorization scales are set to be $\mu = Q$. The LO and NLO singular distributions from SCET perfectly reproduce the full QCD results in the back-to-back limit, which provides a solid check of our factorization formalism.   In the range $\tau \to 1$, the factorization formula does not work well and there are large power corrections, as expected. For small $\tau$ the logarithmic structures in the singular distributions  needed to be resummed to all orders in $\alpha_s$ to obtain stable predictions. 

\begin{figure}
    \centering
    \includegraphics[width=0.49 \textwidth]{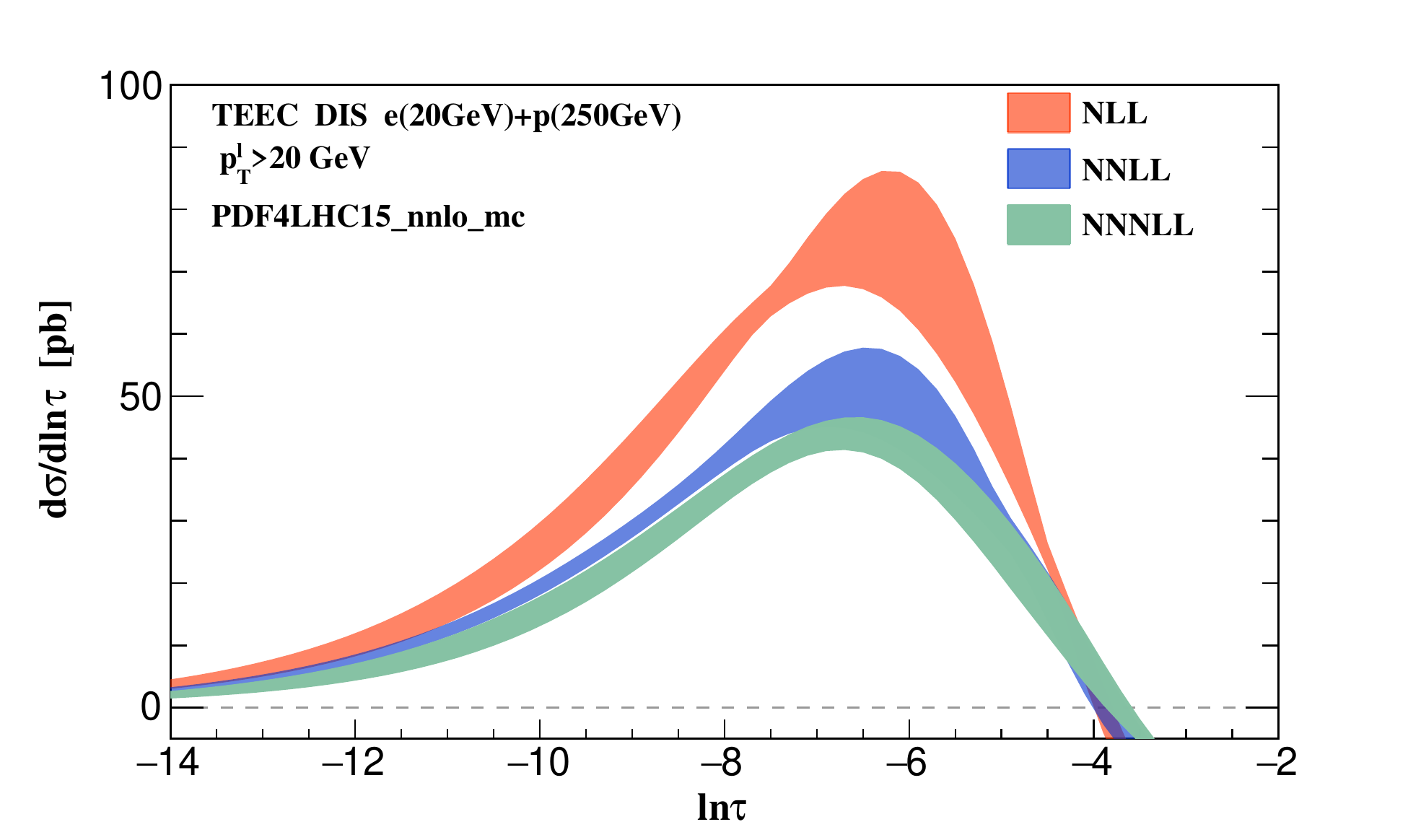}
    \includegraphics[width=0.49 \textwidth]{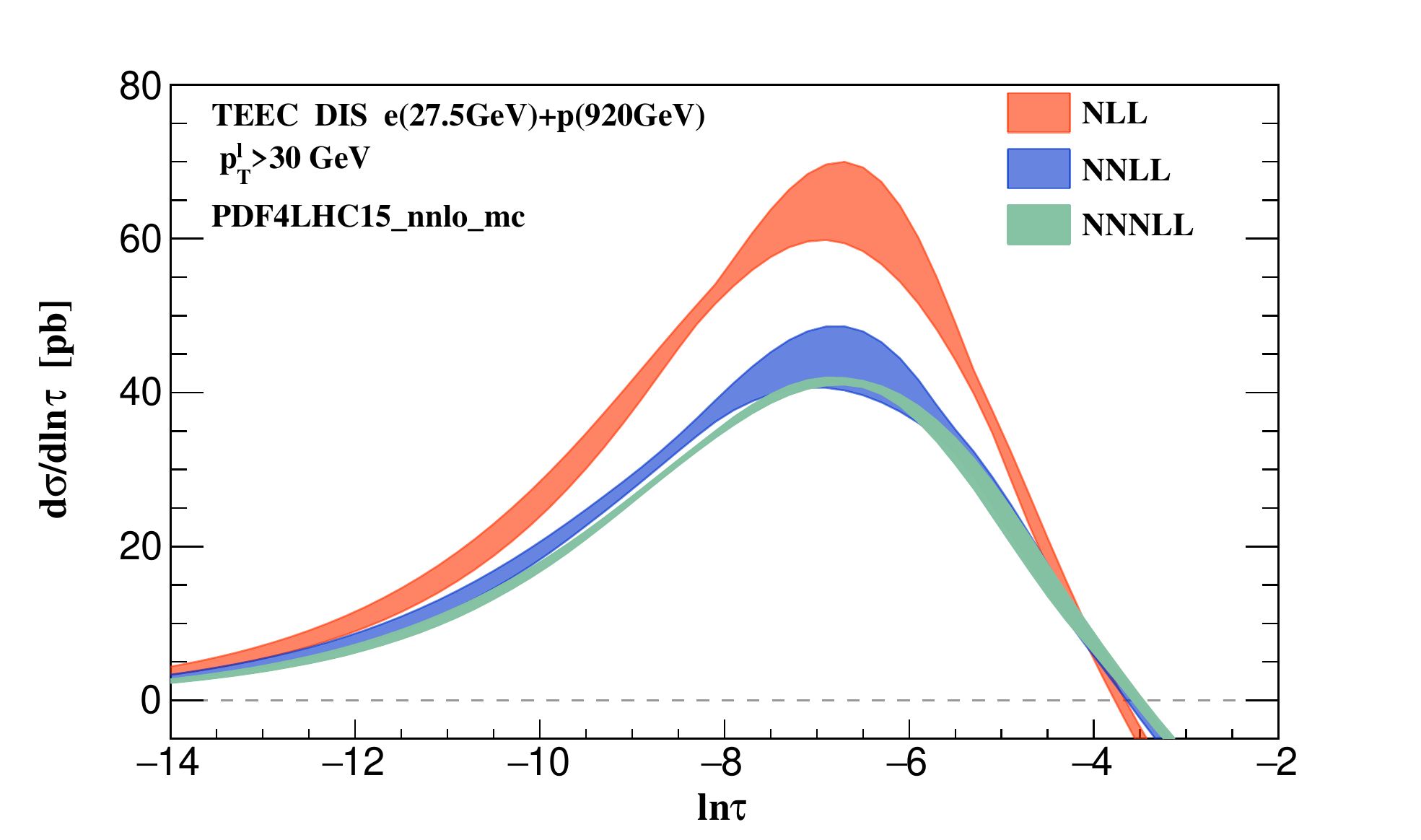}
    \caption{Resummed distributions in the back-to-back limit. The orange, blue, and green bands are the predictions with scale uncertainties at NLL, NNLL and N$^3$LL, respectively.  Left and right panels are for EIC and HERA energies, respectively.}
    \label{fig:resum}
\end{figure}

\subsection{Resummed predictions}

\begin{figure}
    \centering
    \includegraphics[width=0.49 \textwidth]{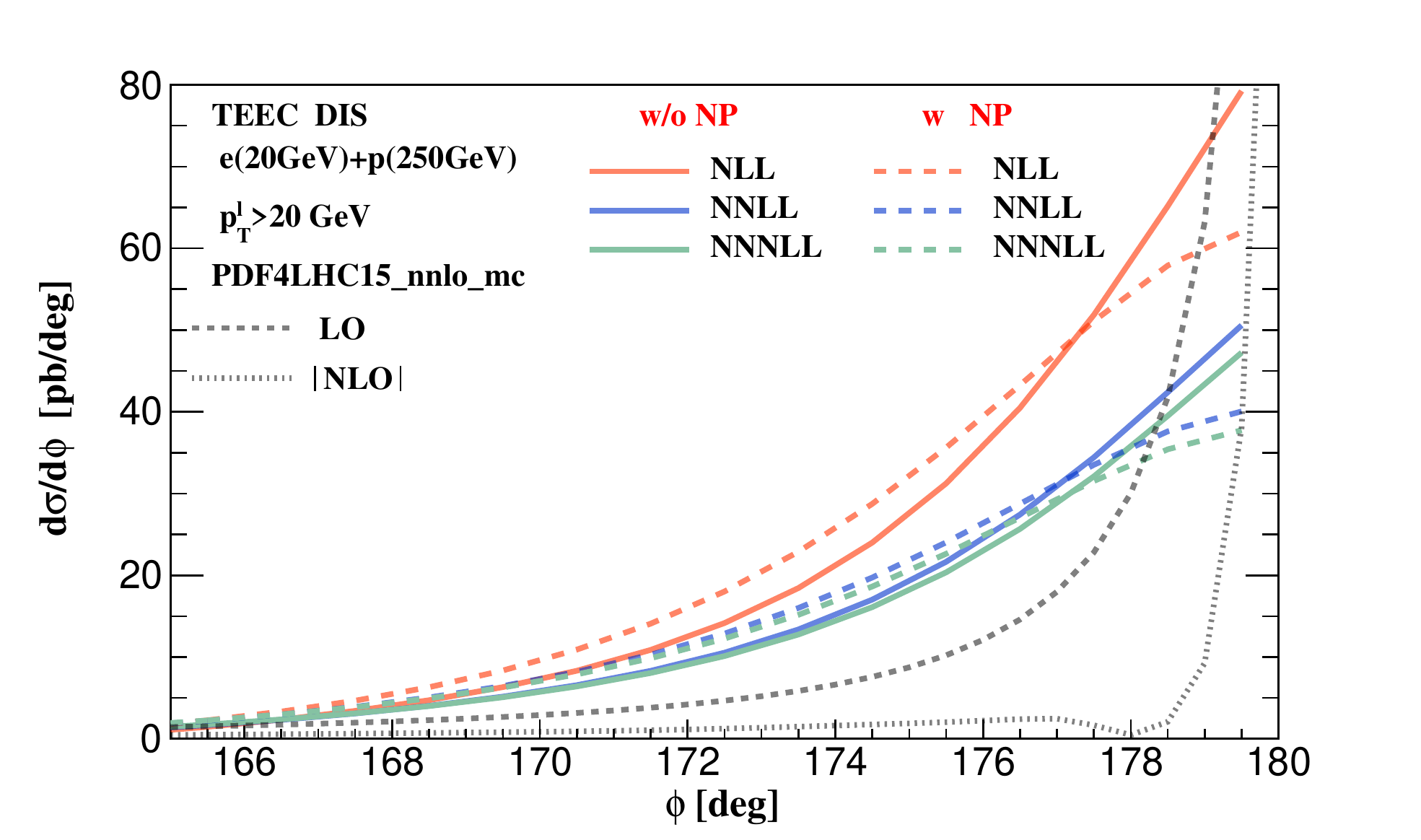}
    \includegraphics[width=0.49 \textwidth]{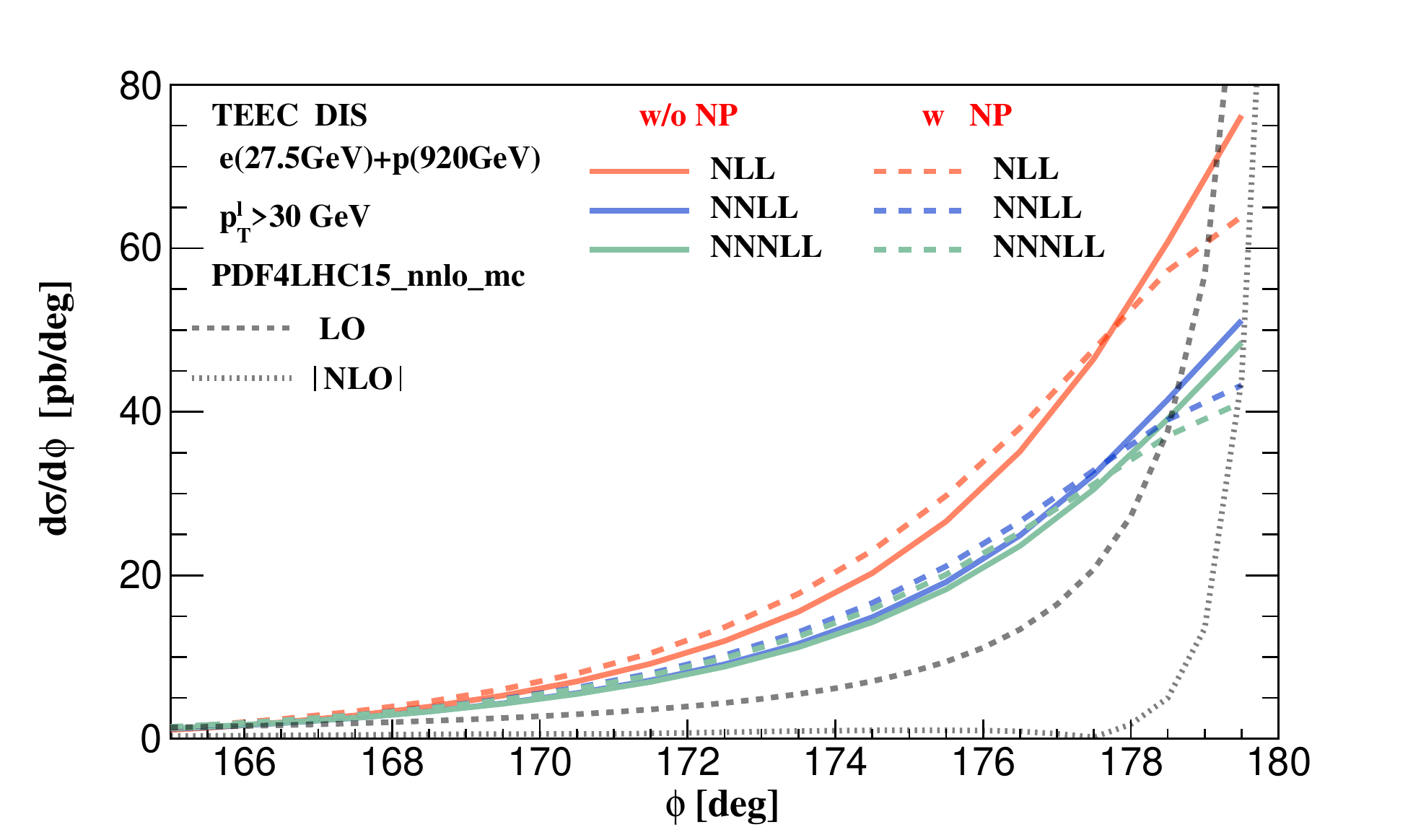}
    \caption{Nonperturbative effects for NLL (orange), NNLL (blue) and N$^3$LL (green) TEEC distributions in DIS. The solid and dashed lines are the predictions without and with nonperturbative effects, respectively. }
    \label{fig:np}
\end{figure}

\begin{figure}
    \centering
    \includegraphics[width=0.49 \textwidth]{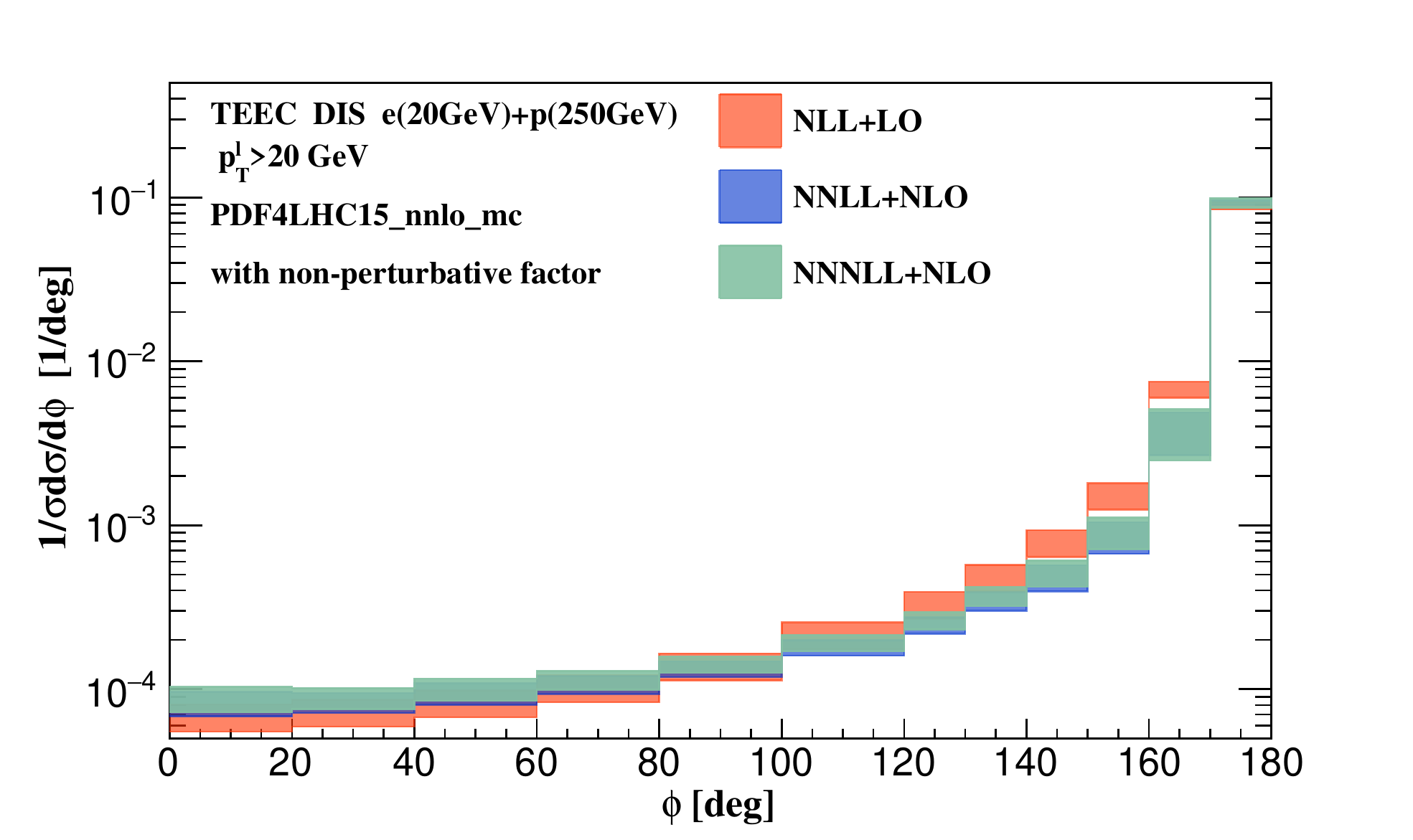}
    \includegraphics[width=0.49 \textwidth]{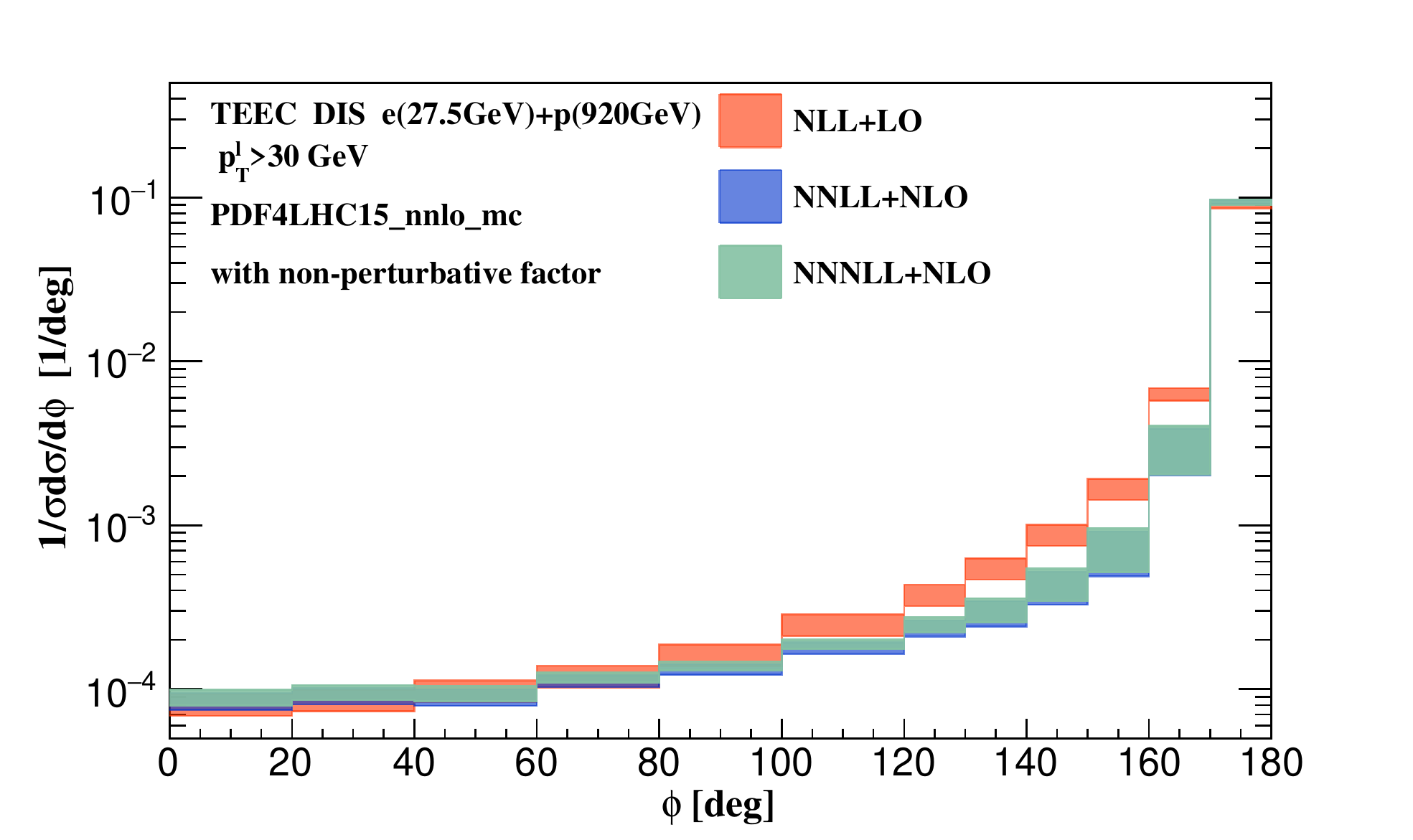}    
    \caption{The TEEC $\phi$ distribution matched with a nonperturbative model. The orange, blue and green bands are the final predictions with scale uncertainties at NLL+LO, NNLL+NLO,  and N$^3$LL+NLO, respectively.}
    \label{fig:phi}
\end{figure}

We set the default scales in our calculation to $\mu_h=\nu=Q$, $\mu_s=\mu=\nu_s=\mu_b$. The scale uncertainties are defined as the quadratic sum over the results when we vary $\mu_h$, $\mu_s$, $\mu$, $\nu_s$ and  $\nu$ independently by a factor of two around their default values.  To avoid Landau pole we use the  $b^*$ prescription~\cite{Collins:1981va}  and define $b^*=b/\sqrt{1+b^2/b^2_{\rm max}}$. We further define $\mu_b = b_0/b^*$ and set $b_{\rm max}=1.5$ GeV$^{-1}$. For completeness, we varied the nonperturbative parameter    $1$~GeV$^{-1} <b_{\rm max}<2$~GeV$^{-1}$. We found that the  dependence on $b_{\rm max}$ of the resummed distributions is very small.

Figure~\ref{fig:resum} presents the resummed predictions at NLL, NNLL,  and N$^3$LL accuracy in the back-to-back limit with scale uncertainties. We find very good perturbative convergence. There is about 30\% suppression in the peak region from NLL to NNLL, while it is about 5-6\% from NNLL to N$^3$LL.  The scale uncertainty of the N$^3$LL result is larger for 141 GeV $ep$ collisions, and is around 12\% near the peak. For 318 GeV collisions the uncertainty is only about 2\%.  The factorization formula gives more accurate predictions at $\sqrt{s} = 318$~GeV due to larger scale hierarchy.  In both cases the uncertainties are significantly improved order-by-order. The resummed distributions turn  negative when $\tau\to1$ where the effective theory becomes invalid.

In general  the nonperturbative  (NP) corrections can be important in the infrared region.  
For EEC in $e^+e^-$ collisions, the work~\cite{Dokshitzer:1999sh} investigated the nonperturbative effects in detail using $\alpha_{\rm eff}$ and a power corrections scheme. 
However, we use a different approach, the $b^*$ prescription, commonly found in hadron collider phenomenology to deal with the Landau pole. 
The nonperturbative effects for TEEC in DIS are related to initial state PDFs, or beam functions, which is  different in comparison with the case of EEC.
 In this work choose the non-perturbative model used in $q_T$ resummation with parameters obtained through fitting  data from refs.~\cite{Su:2014wpa,Prokudin:2015ysa}. 
The nonperturbative effects are included by a multiplicative factor in eq.~(\ref{eq:resum})  
\begin{align} \label{eq:np}
    S_{\rm NP} =\exp\left[-0.106\  b^2 -0.84\ln Q/Q_0 \ln b/b^*  \right] \, , 
\end{align}
with  $Q_0=1.55$ GeV.
Figure~\ref{fig:np} shows the effect of the nonperturbative factor, where the solid and dashed lines are the predictions without and with the NP factor, respectively.
The nonperturbative factor suppresses the cross section for $\phi \sim 180^{o}$ where  low-energy physics is important. The nonperturbative effects are larger in 141 GeV collisions when compared to those in 318 GeV  collisions because the corresponding cross section  in 141 GeV collisions  has a  smaller intrinsic scale. 
Figure~\ref{fig:np} also includes the $d\sigma_{\rm LO}/d\phi$ and  $\big|d\sigma_{\rm NLO}/d\phi \big|$~\footnote{The NLO cross section turns negative when $\phi$ is very close to 180$^o$.}, which are represented by gray dashed and dotted lines. The fixed-order predictions are divergent when $\phi\to 180^{o}$ as expected. Resummation improves these predictions considerably. 

Eq. (\ref{eq:np}) is not the only possible form of nonperturbative corrections. For example, for EEC, the work~\cite{Dokshitzer:1999sh}  considered both a quadratic  and a linear contributions in the impact parameter $b$. The latter can arise from correlations between quarks and soft gluons emitted at the scale of order $\Lambda_{\rm QCD}$ and can give a dominant contribution to the NP effects.   A study along those lines in the future can help us  better understand the interplay between the perturbative and non-perturbative corrections in the back-to-back region.

The results  for the normalized TEEC $\phi$ distributions are shown in Fig.~\ref{fig:phi}, where the nonperturbative factor from eq.~(\ref{eq:np}) is also implemented. The matching region is  chosen to be $160^{o}<\phi<175^{o}$ and for $\phi<160^{o}$ the distributions are generated by fixed-order calculations. The fixed-order predictions are calculated with $\mu_r=\mu_f=\kappa Q$ with $\kappa=(0.5,1,2)$. In the back-to-back limit, the predictions are significantly improved. 
For the second to last bin in Fig.~\ref{fig:phi} 
where $\phi\approx 165^o$ or $\ln\tau \approx -4.1 $, the non-singular contributions are large, which is consistent with Fig.~\ref{fig:fixed-order} and Fig.~\ref{fig:resum}. The scale uncertainties are dominated by the non-singular terms.  The scale uncertainties for the normalized distributions at NNLL+NLO and N$^3$LL+NLO are dominated by the NLO calculations away from the back-to-back region. 
Because of accidental cancellation of the scale dependence for the normalized TEEC distributions at LO, it seems that the NLO QCD corrections do not reduce the scale uncertainties. However, as documented in ref.~\cite{Gehrmann:2019hwf}, the NNLO QCD corrections are expected to reduce the scale dependence of event shape observables significantly.  
Therefore, we expect that matching with NNLO QCD calculations will further reduce the scale uncertainties, but is beyond the scope of this paper.  
The resummation improves the prediction significantly for $\phi\sim 180^o$. There is a small difference for $\phi<160^o$ between NNLL+NLO and N$^{3}$LL+NLO because in a normalized distribution changes in a few bins will affect the distribution in the whole plotted range.  

\section{Conclusion}~\label{sec:concl}

In this paper, we carried out the first study on TEEC in DIS. In the back-to-back limit  the TEEC cross section can be factorized into the product of the hard function, beam function, jet function, and soft function in position space --  closely related to the ordinary TMD physics. We validated the formalism by  comparing our  LO and NLO singular distributions to the full QCD calculations in the back-to-back limit.  The NNLO singular distribution is also provided as a future cross-check of the NNLO cross section for $e+p\to e+$2jet.  The resummed distributions were obtained through solving the RG equations for each component. We found very good perturbative convergence and the scale uncertainties were significantly reduced order-by-order. The nonperturbative effects were assessed using  {\sc\small Pythia} simulations and a nonperturbative model  widely used in $q_T$ resummation.  Importantly, we presented  the first theoretical prediction for $\phi$ distributions at N$^3$LL+NLO accuracy. In the future it will be interesting to consider on the perturbative side matching to an NNLO fixed order calculation~\cite{Gehrmann:2019hwf}. On the nonperturbative side one can explore corrections that arise from quark-gluon correlations and might have a different functional form~\cite{Dokshitzer:1999sh}  than the ones included here.

The EEC/TEEC event shape  observables can be studied in $e^+e^-$, $ep$ and $pp$ collisions, which provides a way to test the universality of QCD factorization in different colliding systems. These observables can also be used to study TMD physics, which is one of the most important goals of the EIC.  Finally, we  remark that TEEC can also be used to shed light on the interaction between partons and  a QCD medium in electron-ion ($eA$) or ion-ion  ($AA$) collisions. Ongoing theoretical and experimental design efforts aim to elucidate the physics opportunities with hadron and jet modification at the EIC and, very importantly, to ensure that  the detectors at this future facility have the capabilities to perform the necessary measurements, see e.g. ref.~\cite{Li:2020sru}. 

As our calculations rely on the SCET framework, a natural choice to address  TEEC in $eA$ collisions  is the extension of the effective theory approach to include the interactions between partons and the background QCD medium mediated by Glauber gluons. Soft collinear effective theory with Glauber gluon interactions has provided a mean to evaluate the contribution in-medium parton  showers~\cite{Ovanesyan:2011xy,Kang:2016ofv} to a variety of observables in reactions with nuclei.  The most recent examples include the modification of jet cross sections and  jet substructure ranging from the jet splitting functions  to the jet 
charge~\cite{Li:2018xuv,Li:2017wwc,Li:2019dre}. To obtain predictions for the TEEC event shape observable in DIS on nuclei will require a computation of the contributions 
from parton branching  in strongly-interacting matter to the terms  in the master factorization formula.  This deserves a separate paper and will be one of our future goals. 

\section*{Acknowledgement}
We  thank F. Yuan and H.X. Zhu for the collaboration at the early stage of the project.  H.T. Li was supported by the Los Alamos National Laboratory LDRD program. I. Vitev was supported by the U.S. Department of Energy under Contract No. DE-AC52-06NA25396 and by the LANL LDRD program. Y.J. Zhu was supported in part by NSFC under contract No. 11975200.

\appendix
\section{Anomalous dimensions} \label{app:ad}

The RG equation of the hard function is 
\begin{align}
    \frac{d}{d\ln\mu} \ln H (Q^2, \mu) \equiv  \Gamma_h = 2 C_F \gamma_{\rm cusp} \ln\frac{Q^2}{\mu^2}+2\gamma_q.
\end{align}
$\gamma_{\rm cusp}$ up to four loops and $\gamma_q$ up to three loops were collected in ref.~\cite{Becher:2019avh} and  references therein. 
The RG equation of the beam/jet function reads
\begin{align}
    \frac{d}{d\ln\mu}\ln \mathcal{G}_i= -C_F \gamma_{\rm cusp}\ln \frac{4 E_i^2}{\nu^2}+\gamma_{G,i} \, , 
\end{align}
where $\mathcal{G}$ represents the beam function $B$ or the jet function $J$.  $E_i$ is the energy of parton $i$.  The RG equation of the soft function is given by
\begin{align}
    \frac{d}{d\ln \mu} \ln S \equiv \Gamma_{s} = -2 C_F \gamma_{\rm cusp} \ln \frac{\nu^2 n_2\cdot n_4}{2 \mu^2} -2 \gamma_{s}~.
\end{align}
The expressions for $\gamma_{s}$ up to three loops can be found in refs.~\cite{Li:2016ctv, Moult:2018jzp}.
Additionally we have $\gamma_{B,q}=\gamma_{J,q}$, which  can be derived from $2\gamma_q +\gamma_{B,q}+\gamma_{J,q}-2 \gamma_{s}=0$ according to the scale invariance of the cross sections.  Note that 
\begin{align}
    2 \ln \frac{Q^2}{\mu^2} - \ln \frac{4 E_2^2}{\nu^2} - \ln \frac{4 E_4^2}{\nu^2} - 2\ln\frac{\nu^2 n_2 \cdot n_4}{2\mu^2} = 0 \, , 
\end{align}
with $Q^2 = 4 E_2 E_4  \frac{n_2 \cdot n_4 }{2}$~.

The rapidity evolution equation of the beam/jet function reads
\begin{align}
    \frac{d}{d\ln\nu}\ln  \mathcal{G}_{i} =  C_F\left[\int_{b_0^2/b^2}^{\mu^2} \frac{d\bar{\mu}^2}{\bar{\mu}^2} \gamma_{\rm cusp}(\bar{\mu}) -\gamma_r(b_0^2/b^2) \right]  ]\, . 
\end{align}
Similarly, the rapidity evolution equation of the soft function is given by
\begin{align}
    \frac{d}{d\ln\nu} \ln S =  2 C_F\left[-\int_{b_0^2/b^2}^{\mu^2} \frac{d\bar{\mu}^2}{\bar{\mu}^2} \gamma_{\rm cusp}(\bar{\mu}) +\gamma_r(b_0^2/b^2) \right] \, , 
\end{align}
where $\gamma_r$ can be found in refs.~\cite{Li:2016ctv, Moult:2018jzp}.  The cross section is independent of $\nu$, which leads to the constraint 
\begin{align}
     \frac{d}{d\ln\nu}\ln  B_{q}+ \frac{d}{d\ln\nu}\ln  J_{q}+ \frac{d}{d\ln\nu}\ln  S =0~. 
\end{align}

\bibliographystyle{JHEP}
\bibliography{bib}
\end{document}